\pdfoutput=1 
\documentclass[sigconf]{acmart}
\usepackage{algorithm2e}
\usepackage{graphicx}
\usepackage{subcaption}
\usepackage{multirow}
\usepackage{listings}
\acmConference[preprint]{under review}{}{online}
\begin{document}

\title{foREST: A Tree-based Approach for Fuzzing RESTful APIs}

\author{Jiaxian Lin$^\dagger$, Tianyu Li$^\dagger$, Yang Chen$^\dagger$, Guangsheng Wei$^\ddagger$, Jiadong Lin$^\ddagger$, Sen Zhang$^\ddagger$, Hui Xu$^\dagger$}
\authornote{Corresponding author.}
\affiliation{
  \institution{$^\dagger$School of Computer Science, Fudan University}
  \country{China}
}
\affiliation{
  \institution{$^\ddagger$Huawei Cloud BU} 
  \country{China}
}

\begin{abstract}
Representational state transfer (REST) is a widely employed architecture by web applications and cloud. Users can invoke such services according to the specification of their application interfaces, namely RESTful APIs. Existing approaches for fuzzing RESTful APIs are generally based on classic API-dependency graphs. However, such dependencies are inefficient for REST services due to the explosion of dependencies among APIs. In this paper, we propose a novel tree-based approach that can better capture the essential dependencies and largely improve the efficiency of RESTful API fuzzing. In particular, the hierarchical information of the endpoints across multiple APIs enables us to construct an API tree, and the relationships of tree nodes can indicate the priority of resource dependencies, \textit{e.g.,} it's more likely that a node depends on its parent node rather than its offspring or siblings. In the evaluation part, we first confirm that such a tree-based approach is more efficient than traditional graph-based approaches. We then apply our tool to fuzz two real-world RESTful services and compare the performance with two state-of-the-art tools, EvoMaster and RESTler. Our results show that foREST can improve the code coverage in all experiments, ranging from 11.5\% to 82.5\%. Besides, our tool finds 11 new bugs previously unknown.
\end{abstract}

\maketitle

\section{Introduction}
Nowadays, REST becomes a defacto architectural standard for web applications and cloud~\cite{gamez2017analysis}. The architecture views each REST service as a storage system that can process data operation requests from clients and make responses. Similar to the classic CRUD (create, read, update, and delete) operations on storage systems, REST also defines four corresponding methods, GET, POST, PUT, and DELETE. The communication between REST clients and services is based on HTTP requests and responses, which brings the architecture excellent interoperability in nature. Due to such advantages, REST has been widely employed by industries. For example, both Microsoft Azure\footnote{https://docs.microsoft.com/en-us/rest/api/azure/} and Google Cloud\footnote{https://cloud.google.com/apis} adopt REST, and users can access their provided services via RESTful APIs.

To assure the quality of REST software, testing is an essential approach. While traditional unit test or integration test requires much human effort in writing test cases, fuzzing is a favorable testing approach that can be fully automated. Given the API specification of a target REST service as input, a fuzzing engine can generate test cases automatically and exercise each test case with the service. The main challenge of fuzzing lies in how to solve the dependencies among RESTful APIs in order to generate valid and diversified test cases. Existing work (\textit{e.g.,} RESTler~\cite{atlidakis2019restler} and RestTestGen\cite{viglianisi2020resttestgen}) on this direction mainly employs a straightforward API dependency graph. By traversing the graph via breadth-first search (BFS) or topological sort, a fuzzing engine should be able to generate valid test cases. However, the efficiency of such approaches is limited due to the explosion of edges on the dependency graph. As a result, there could be a huge number of possible paths for reaching an API. Furthermore, since some dependencies are inaccurate (discussed in Section~\ref{sec:limitation}), it is hard for such a tool to arbitrarily choose only one valid path and abandon the rest.

In this paper, we propose a novel tree-based approach for fuzzing RESTful APIs. Our approach can capture the essential dependencies among APIs via a tree structure and therefore improve the dependency complexity from quadratic (traversing a graph) to linear (traversing a tree) with respect to the number of APIs. Note that each RESTful API is uniquely defined with an endpoint and a method (GET, POST, PUT, or DELETE). Rather than modeling the dependencies of APIs based on their required parameters and response values, we extract API hierarchies according to the endpoint of each API. Our idea is inspired by the fact that the required parameters or resources for visiting a child node on an endpoint tree have a high chance to be fulfilled once its parent node has been visited. We can therefore pre-order traverse the tree to generate valid test cases. Moreover, the relationships among tree nodes can indicate the priority for acquiring the dependent resource. For example, it's more likely that a node depends on its parent node than its descendent or siblings, although there could be dependencies among all of them according to the traditional dependency graph. In this way, our approach can improve the possibility of finding useful resources for generating valid requests and improve the efficiency of fuzzing. 

To elaborate, our approach parses RESTful API specifications in Swagger or OpenAPI standard~\cite{openapi2017openapi} and constructs endpoint trees for these APIs. To this end, it splits each endpoint URL (uniform resource locator) into several tokens by the slash symbol. Each token is a node on an endpoint tree, and we connect these nodes with edges such that an endpoint URL can be reconstructed as a path from the root to a descendent node. Meanwhile, each node may have several attributes indicating the supported methods of the endpoint, and a buffer to stores the resources of the node. When generating test cases for each node, we employ a template-based method that specifies an ad hoc order of HTTP request methods, \textit{e.g.,} GET, then several POST, followed by PUT and DELETE. Since PUT and DELETE may invalidate the created resource, we should execute POST more times than DELETE in order to leave valid resources for the decedent nodes to use. Note that there could be multiple trees for an application. Our approach should also work in this scenario by visiting each tree iteratively. The complexity is still linear to the number of APIs.

We have implemented a prototype for the tree-based approach, namely foREST. Besides the basic tree-based model for guiding request sequence generation, foREST has also realized the associated feature for resource management and retrieval. We have conducted a set of comparison experiments to study the efficiency of our approach. Our experimental result firstly verifies that such a tree-based approach is more efficient than traditional graph-based approaches in practice, including BFS and topological sort. Furthermore, we compare foREST with two state-of-the-art tools for RESTful API fuzzing, \textit{i.e.,} RESTler~\cite{atlidakis2019restler} and EvoMaster~\cite{EvoMaster}. We fuzz WordPress and GitLab with these three tools, and each fuzzing experiment lasts for six hours. Results show that foREST can achieve the best code coverage in all experiments and has exceeded the second best one (EvoMaster) by 11.5\% to 82.5\% in different experiments. Besides, foREST has found 11 new bugs previously unknown, while RESTler and EvoMaster have found none.

In short, this article contains several major research contributions as follows.

\begin{itemize}
\item It proposes a novel tree-based RESTful API fuzzing approach that is more efficient than traditional graph-based approaches. To our best knowledge, it is the first systematic work in this direction. Our approach subtly models the relations of APIs with a tree structure which not only reduces the complexity of API dependencies but also captures the priority of resource dependencies.

\item We have implemented a prototype tool, foREST, and released it as open source on GitHub\footnote{The link will be released soon.}. Our fuzzing experimental results with real-world REST services show that foREST has achieved better performance than state-of-the-art tools. We believe our approach and the tool would be useful to advance the development of the community on the problem of RESTful API fuzzing.

\end{itemize}

The rest of the article is organized as follows. Section~\ref{sec:problem} firstly introduces the problem of RESTful API fuzzing and discusses the challenges of the problem. Section~\ref{sec:approach} then presents a motivating example for employing a tree-based approach and demonstrating our methodology. Section~\ref{sec:evaluation} evaluates the performance of our approach. Section~\ref{sec:related} discusses related work, and finally Section~\ref{sec:conclusion} concludes the paper.

\begin{figure}[t]
\centering
\begin{subfigure}{0.5\textwidth}
\includegraphics[width=\linewidth]{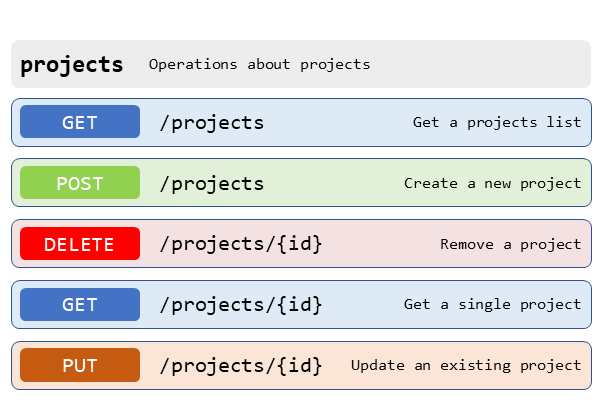}
\caption{Five sample APIs.}
\label{fig:gitlab-project}
\end{subfigure}
\begin{subfigure}{0.5\textwidth}
\includegraphics[width=\linewidth]{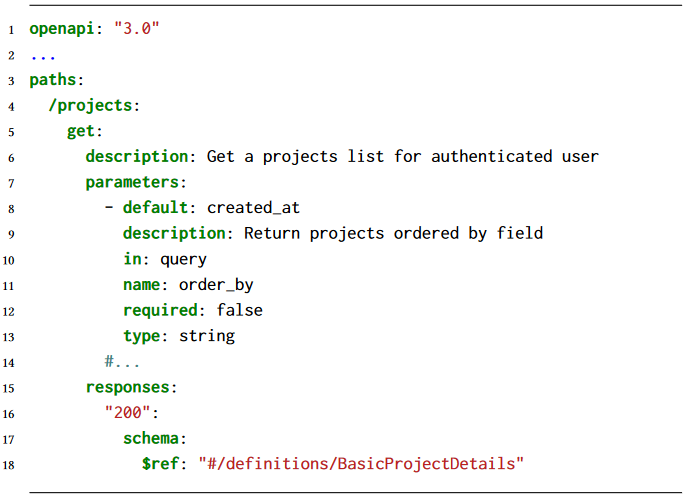}
\caption{Definition of the API \text{GET /projects} in OpenAPI 3.0.}
\label{fig:gitlab-prject-yaml}
\end{subfigure}
\caption{Sample APIs of the GitLab project.}
\label{fig:gitlab-apis}
\end{figure}

\begin{figure*}[ht]
\centering
\begin{subfigure}{0.35\textwidth}
\includegraphics[width=\linewidth]{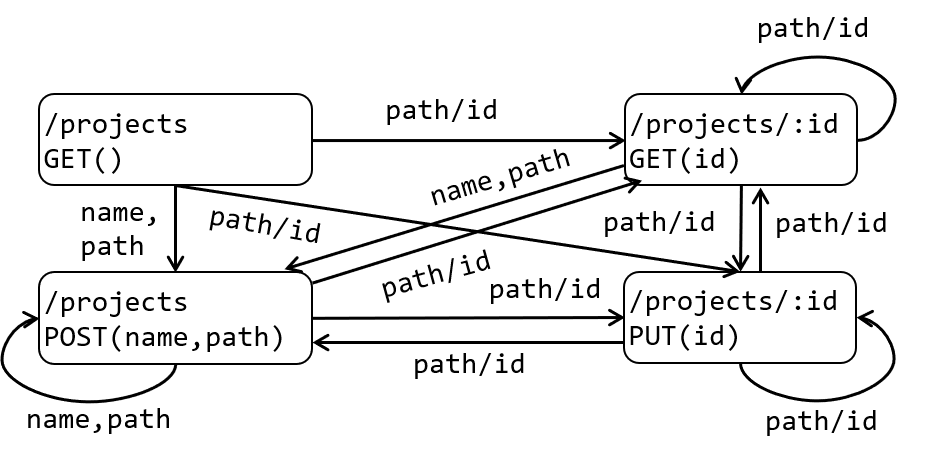}
\caption{Dependency graph of four APIs.}
\label{fig:graph-project}
\end{subfigure}
\hspace{0.1cm}
\begin{subfigure}{0.63\textwidth}
\includegraphics[width=\linewidth]{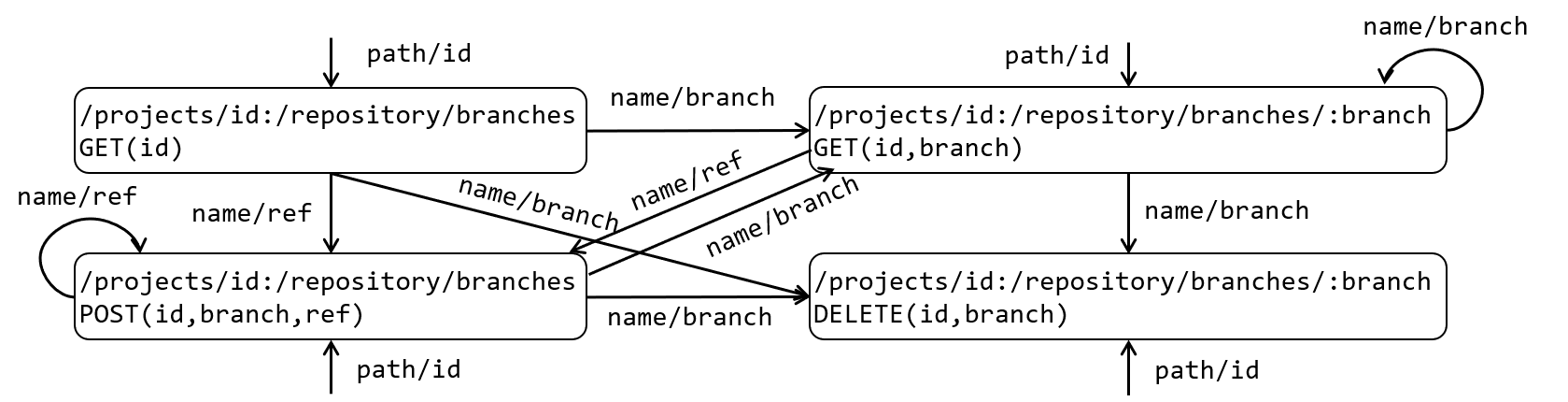}
\caption{Dependency graph when considering more APIs. We mainly provide several local dependencies of a cropped graph for clarify.}
\label{fig:graph-branch}
\end{subfigure}
\caption{Representing sample GitLab APIs with traditional API dependency graphs. An arrow $A\xrightarrow[]{\text{resp/req}}B$ indicate B depends on A, and the required parameter name in A's response is \texttt{resp}, and in B's request is \texttt{req}. For instance, \texttt{path/id} means the required parameter is \texttt{id}, but it uses another name \texttt{path} in the response.}
\label{fig:graph}
\end{figure*}
\section{Problem and Challenges}\label{sec:problem}
\subsection{Problem of RESTful API Fuzzing}
RESTful API is a common way for clients to interact with services. The latest standard for defining such APIs is OpenAPI 3.0~\cite{openapi2017openapi}. In general, an RESTful API specification should provide both the format of messages for clients to launch valid requests and the guidance for interpreting message responses. A request message is composed of the following elements: 

\begin{itemize}
\item \textit{Endpoint}. 
An endpoint is a path string, which indicates the specific URL of the API. Besides the domain name or IP address within an URL, each path may consist of several tokens or arguments separated by ``/''. While a token is a common substring, an argument is a substring started with a colon or wrapped with a bracelet. For example, \texttt{/token/:arg} and \texttt{/token/\{arg\}} implies the same path with one argument \texttt{arg}.

\item \textit{Method}. RESTful API is based on HTTP, and it supports several HTTP methods, including POST for creating data, GET for retrieving data, PUT for updating data, PATCH (rarely used in practice) for a partial update, and DELETE for removing data.

\item \textit{Parameter}. Besides the arguments inline with the endpoint URL, an RESTful API can also accept parameters from the HTTP content in the form of key-value pairs. Depending on the design of the API, such parameters could either be required or optional. 
\end{itemize}

Each RESTful API is uniquely defined by its path and method. In general, a path may supports multiple methods, and they are treated as different APIs. For example, Figure~\ref{fig:gitlab-project} contains five different APIs of GitLab, which belong to two paths, \texttt{/projects} and \texttt{/projects/\{id\}}. Figure~\ref{fig:gitlab-prject-yaml} demonstrates the detailed specification of one API, \texttt{GET} \texttt{/projects}. The API specification includes both the request and response message format. For requests, user can specify a string value for the optional parameter ``\texttt{order\_by}''. Otherwise, the default value for ``\texttt{order\_by}'' would be ``\texttt{created\_at}''. For responses, a response code 200 indicates that the request is properly received by the server and the response message can be decoded according to the specification of \texttt{\#\/definitions/BasicProjectDetails}. 

The problem of RESTful API fuzzing lies in how to generate test cases given the API specification of an RESTful service and practice them with the service. Such fuzzing solutions are often evaluated against how many APIs a fuzzer can activate or how many lines of code it can cover with a given time or test case budget. In order to generate valid requests for a target API and exercise more lines of code, the fuzzer should be able to supply valid parameter values automatically by solving the dependencies of the API. Next, we discuss several main challenges underlying the problem.

\subsection{Challenges for Fuzzing RESTful APIs}\label{sec:challenge}

Real-world RESTful services generally contain data dependencies among APIs. For example, Figure~\ref{fig:graph-project} is a dependency graph with detailed relationships among four APIs of Figure~\ref{fig:gitlab-apis}. To generate a valid test case for the API \texttt{GET} \texttt{/projects/:id}, the test case should pass a legal parameter \texttt{id}, which can be obtained from the response message of another API \texttt{GET} \texttt{/projects} or \texttt{POST} \texttt{/projects}. Therefore, the main challenge of RESTful API fuzzing lies in how to model and solve these data dependencies in order to generate valid test cases for each API.

Note that the problem is different from existing fuzz target generation problems for library APIs, such as~\cite{babic2019fudge,jiang2021rulf}. A library API generally takes one or several parameters and returns one value. In order to generate a valid API call as the fuzz target, the API parameter types should match the API signature. Most of these types are primitive ones (such as \texttt{int} or \texttt{char}) or abstract data types that can be returned by other APIs of the library. Therefore, there are also data dependencies among the parameters and return values of different APIs. Below, we discuss three typical differences between traditional fuzz target generation problems and RESTful API fuzzing.

\textit{M to N:} Compared to a library API that returns only one value (\textit{i.e.,} $f(x_1,...,x_m)\to y$), the response message of an RESTful API generally contains multiple values (\textit{i.e.,} $f(x_1,...,x_m)\to (y_1,...,y_n)$). Therefore, the dependencies should be more complicated, \textit{i.e.,} the parameter of an RESTful API depends on one particular field of the response from another API. Note that a library API may also return multiple values as a tuple, but it generally also contains another straightforward constructor for the required type.

\textit{Arbitrary matching:} the dependencies of RESTful APIs are based on the names of values instead of the function signatures (\textit{i.e.,} $t_{x_1},...,t_{x_n}\to t_{y}$) for library APIs. Developers can arbitrarily define their names, and there is no strict rules similar to type checking. As a result, developers may choose different names among different APIs for the same resource, or using the same name for different resources. Such issues are very popular in real-world REST applications, leading to another challenge for fuzzing RESTful APIs.

\textit{Resource dependency:} RESTful applications are generally stateful, and the API dependencies should also consider the state or resources maintained by the application. Otherwise, the generated data dependencies by simply matching the resource name could be fake. In particular, only POST or PUT creates new resources. The response of GET or DELETE may also contain resources, but these resources are not newly created.

\begin{figure*}[th]
\centering
\includegraphics[width=0.99\textwidth]{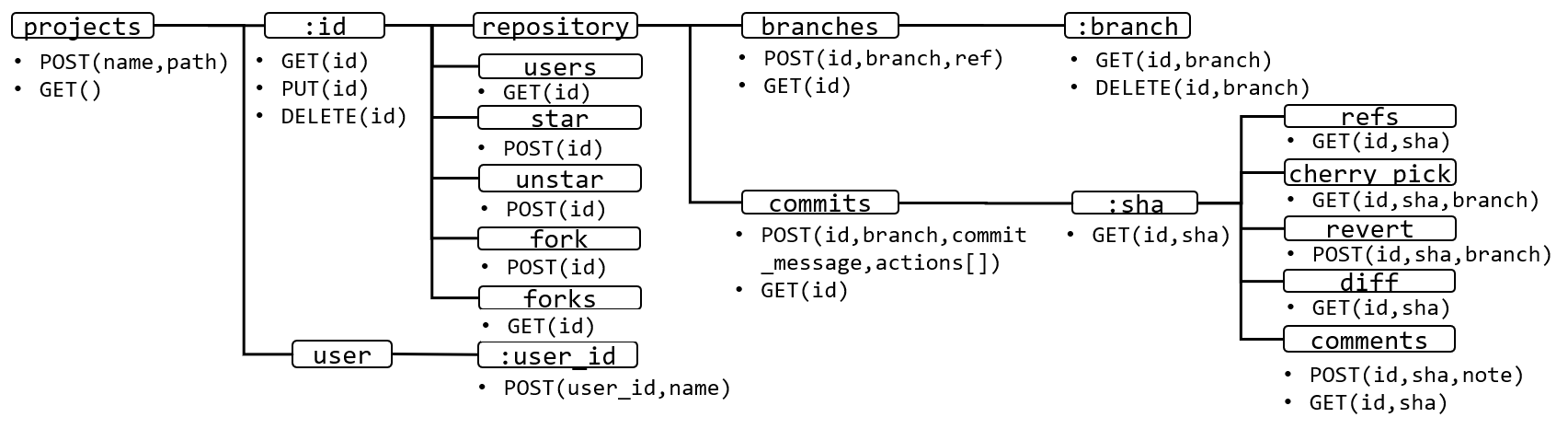}
\caption{Representing sample GitLab APIs as a tree. Each node may support multiple HTTP methods, and the endpoint URL can be reconstructed as a path from the root to each node.}
\label{fig:tree_hierarchy}
\end{figure*}

\section{Tree-based Approach}\label{sec:approach}
Since there are resource dependencies among APIs, we adopt a tree-based approach to model and solve such dependencies. This section first presents a motivating example showing the advantage of our approach, followed by our detailed methodology.

\subsection{Motivating Example}
\subsubsection{Limitations of API Dependency Graph}\label{sec:limitation}

We cannot adopt traditional signature-based API dependency because it is inaccurate for capturing the resource dependencies among RESTful APIs. As a result, fuzzing RESTful APIs via API dependency graph traversal would encounter severe issues.

We demonstrate our point with Figure~\ref{fig:graph}.  Figure~\ref{fig:graph-project} constructs a sample API dependency graph based on the request and response message format specified in the API definition of the GitLab project. We have manually matched all arbitrary names that denote the same resource. For example, \texttt{path} and {id} are two identifiers used in the response and request for the same thing, and we can denote the dependency edge as \texttt{path/id}. In order to traverse the graph, we should follow an order that can solve such data dependencies, \textit{e.g.,} via BFS or topological sort. Therefore, we should invoke \texttt{GET} \texttt{/projects} first because the API has no dependencies. Then we can extract the values of \texttt{name} and \texttt{path} from its response as the parameter values for composing requests for the rest three APIs. The order apparently works but suffers several essential problems in practice. 

\begin{itemize}
\item \textit{Dense graph}: Since an RESTful API response generally contains multiple values, the dependency graph could be highly dense, \textit{i.e.,} many edges among API nodes. For example, all the responses of the four APIs in Figure~\ref{fig:graph-project} contain a \texttt{path} field, and three APIs require the value as the parameter value of \textit{id}. Besides, many other GitLab APIs (such as those in Figure~\ref{fig:graph-branch}) also require \texttt{path/id}. Traversing a dense graph is very challenging due to the path explosion problem and an efficient algorithm may neglect some important routes that should be explored.

\item \textit{Fake producer}: We should not treat all API dependencies equally because some dependencies are fake. For example, if no project has been created yet, the response of \texttt{GET} \texttt{/projects} could be null. In other words, \texttt{GET} \texttt{/projects} is not the real producer of the resources \texttt{name} and \texttt{path} required by other APIs.

\item \textit{Weak dependency}: There is another type of dependencies should be treated differently. For example, \texttt{POST} \texttt{/projects} depends on \texttt{GET} \texttt{/projects} for the parameters of \texttt{name} and {path}. However, such resources are conflicting because these two parameters should be uniquely initiated by users. One cannot create a project with an existing name or path. Imposing such a dependency is mainly useful for reaching the code for duplication check.

\end{itemize}

\begin{figure*}[th]
\centering
\includegraphics[width=0.99\textwidth]{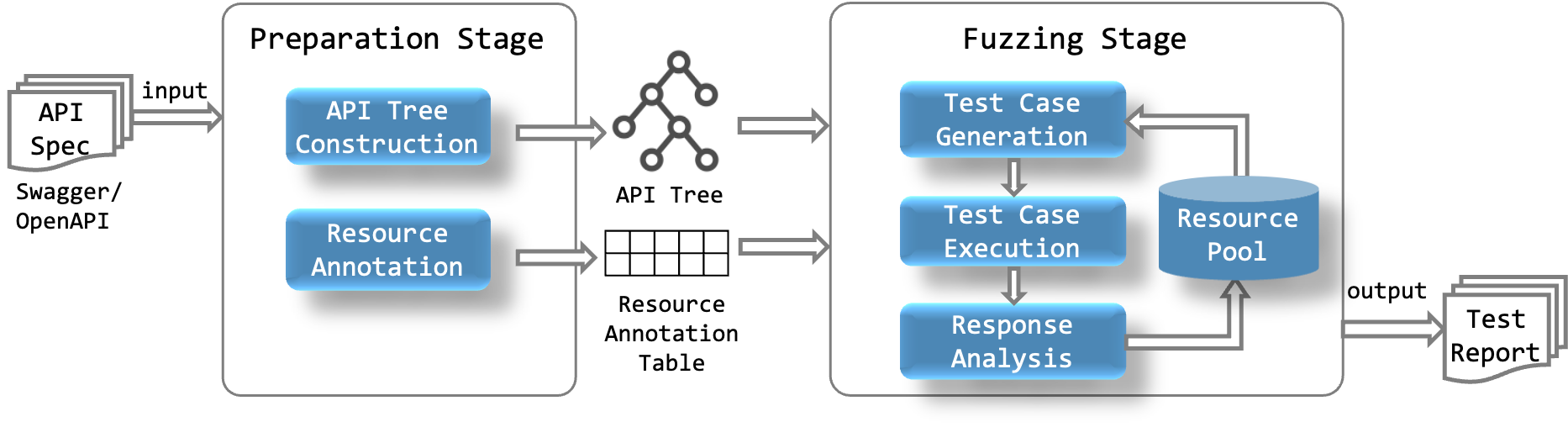}
\caption{Overall framework of foREST.}
\label{fig:framework}
\end{figure*}

\subsubsection{Hierarchy of RESTful APIs}\label{sec:api-hierachy}

We observe that the paths of well-designed RESTful APIs generally form a tree structure (or forest) in nature.  Figure~\ref{fig:tree_hierarchy} contains dozens of GitLab APIs, which is a super set of the APIs in Figure~\ref{fig:graph}. Each node on the graph indicates one component of the endpoint URL separated by ``/''. Most nodes have one or several method attributes beneath them, denoting the request methods supported by an endpoint. The URL of each API can be reconstructed as a path from the root node (\texttt{projects}) to another node with method attributes. For example, \texttt{/projects}, \texttt{/projects/:id}, and \texttt{/projects/users} are all valid URLs. Based on the content of a node, we further divide them into two types: \textit{token node} which contains a fixed path string, and \textit{parameter node} which has a parameter to be specified. 

Such API trees subtly capture the resource dependencies among APIs. In particular, if an API needs several parameters, it is likely that these parameter values can be produced by another endpoint along the same path but has shorter URL. For instance, the API \texttt{GET} \texttt{/projects/:id} requires one parameter \texttt{id}, which can be produced by \texttt{POST} \texttt{/projects}; \texttt{DELETE} \texttt{/projects/:id/repositories/branches/:branch} requires two parameters \texttt{id} and \texttt{branch}, which can be produced by \texttt{POST} \texttt{/projects} and \texttt{POST} \texttt{/projects/:id/repositories/branches} correspondingly. Such dependencies are based on the production and use (or consumption) of resources, which are essential for RESTful API testing.

\subsection{Methodology}
Figure~\ref{fig:framework} demonstrates the overall framework of our approach. There are two stages: a preparation stage and a fuzzing stage. In the preparation stage, we organize all the APIs into trees. Besides, we need to annotate some arbitrary resource names with unique identifiers. In the fuzzing stage, we generate test cases by traversing the API trees, and then execute each of them with the target REST service. Meanwhile, we analyze the response messages and extract useful resources for reuse based on a resource pool. Next, we discuss the detailed design of several key procedures.

\subsubsection{API Tree Construction}
This step parses the specification document of RESTful APIs and organize them into trees. As discussed in Section~\ref{sec:api-hierachy}, an API tree can be defined as $T(V,E)$, where each node $v_i$ is a component of an URL (without the domain name or IP address) separated by ``/'', and an edge $e_{i,j}$ that connects $v_i$ and $v_j$ exists if and only if $v_i/v_j$ appears in an URL. Each node $v_i$ may have one or several method attributes, denoting the API methods supported by the URL from the root node $v_0$ to $v_i$, \textit{i.e.,} the path can be restored from the tree as $/v_0/.../v_i$.

\subsubsection{Resource Annotation and Fuzzy Matching}~\label{sec:fuzzymatching}
We design a resource annotation table to deal with those arbitrary names of resources. The index of the table are unified resource identities, and each resource may have several arbitrary names in the specification. Our resource annotation table is in nature an array of name sets, \textit{i.e.,} $$\bigcup\limits_{i=1}^{n} \{R_{i}|R_{i}=\{name_1,...,name_m\}\}$$.

Since filling in the table manually by developers would be labour intensive, we employ another fuzzy matching strategy to automatically recognize some potential name pairs commonly used in RESTful APIs. Our matching rule ignores the capitalization of letters, spaces and underlines between letters, \textit{etc}. In this way, developers only need to annotate a limited number of names that are very different. Besides, we also employ a scoring mechanism to automatically filter falsely matched pairs, \textit{i.e.,} if the response code based on a fuzzy pair is 4XX, we give the pair a low score and tend to abandon it in the next round. 

\subsubsection{Resource Pool}
We employ a resource pool to buffer and reuse dependent resources. The structure of our resource pool is consistent with the hierarchical relationship of APIs. In other words, we create a sub pool for each token node of the API tree. Since only particular resource combinations could satisfy the business requirement of the service, we do not record each resource separately but document them together as a tuple if they occur in the same successful request or response. Note that it is unnecessary to create a sub pool for the parameter node which may have different values, because such node generally indicates a specific resource that has already been included in its parent token node. For example, the token node \texttt{/projects} contains all project resources, and \texttt{/projects/{id}} contains only one particular project indicated by the project \texttt{id}.

Besides, the hierarchy of our resource pool can also indicate the priority of dependency. When generating a request for an API, we first search resources from the current node for the required parameters, and then its parent node, followed by other ancestor nodes.

\SetKwComment{Comment}{/* }{ */}
\begin{algorithm}[t]
\caption{Basic test case generation algorithm for required parameters only.}
\label{alg:fuzzing}
\KwData{$deptrees$ /*API hierarchical trees*/}
\KwData{$annotable$ /*Resource annotation table*/} 
\KwData{$respool$ /*Resource pool, initiated as null*/} 
\For {$t \in apitrees$} {
   $nodeseq \gets DFS(t)$ /*get a node sequence in DFS order*/\\
   \For{$n \in nodeseq$} {
       $url \gets GetEndpoint(n,t) $ \\
       $methods \gets GetMethod(n,t)$ /*get the methods of an endpoint and save them in a predefined order*/ \\ 
       \For{$m \in methods$}{ 
         $pars \gets GetRequiredParam(url,m)$\\
         $pairs \gets null$ /*for rating purpose*/\\
         \For{$p \in pars$}{
            $rid \gets Search(p,annotable)$ /*search the unique id from the res. annotation table*/\\
            $rid \gets FuzzyMatching(rid)$ /*return a res. id along the tree backward or none.*/\\
            $pairs.add(p,rid)$
            $p.v = Retrieve(rid,respool)$ /*find a value from the resource pool randomly */
         }
         $resp = Execute(url, m, pars)$\\
         \If{$resp.code==2XX || 3XX$}{
            $res = Extract(resp.msg)$\\
            $respool.add(res)$\\
            $IncreaseMatchingScore(pairs)$\\
        }
        \If{$resp.code==4XX$}{
            $DecreaseMatchingScore(pairs)$
        }
        \If{$resp.code==5XX$}{
            $ReportBug()$\\
        }
      }
   }
}
\end{algorithm}

\begin{figure}[t]
\centering
\begin{subfigure}{0.49\textwidth}
\includegraphics[width=\linewidth]{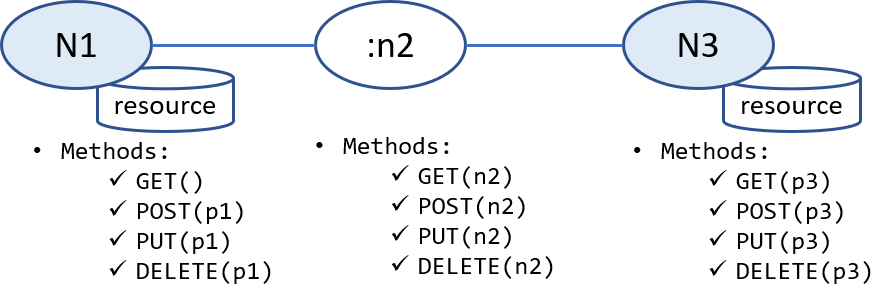}
\caption{A toy API tree example.}
\label{fig:alg-example}
\end{subfigure}
\begin{subfigure}{0.49\textwidth}
\includegraphics[width=\linewidth]{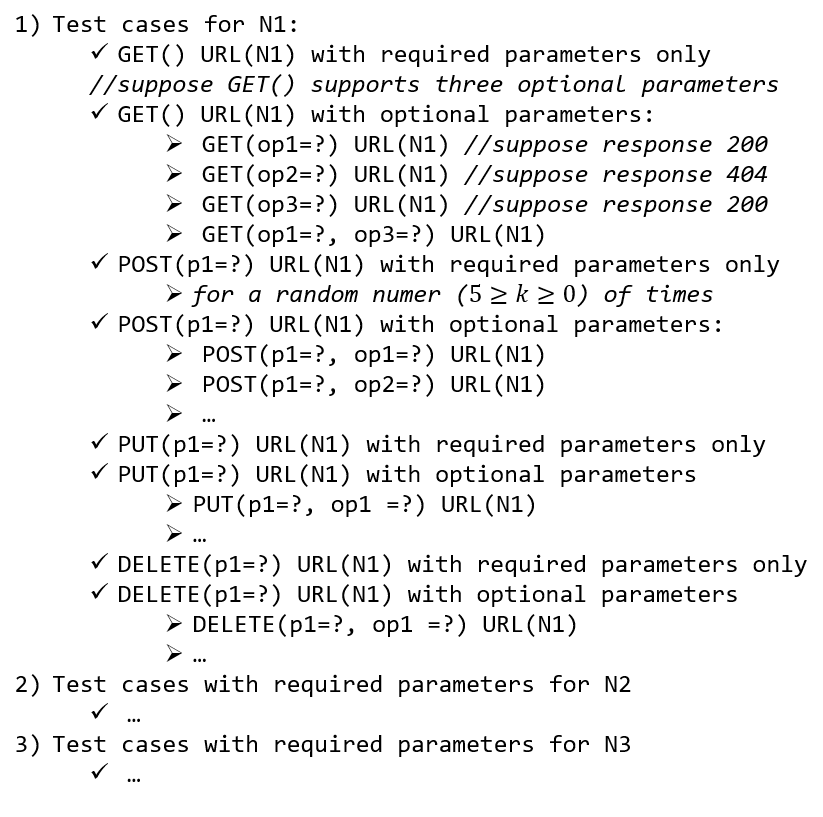}
\caption{Sequences of test cases generated.}
\label{fig:alg-example-seq}
\end{subfigure}
\caption{Demonstrating the test cases generated by our approach considering both required parameters and optional parameters with a toy API tree.}
\label{fig:alg-demo}
\end{figure}

\subsubsection{Test Case Generation}
The tree structure provides us with an essential reference for generating valid test cases. Our basic idea is that we can traverse each tree via either a depth-first or bread-first order and generate valid test cases for each node. Since a node may support multiple HTTP methods, we adopt template-based order for exercising the methods of each node, where a template is a heuristic order of HTTP methods, \textit{e.g.,} \texttt{GET}, and then $k$ times of \texttt{POST}, and then \texttt{PUT}, followed by \texttt{DELETE}, where $k$ is a small random integer generated during runtime for the fuzzing purpose. The order is based on the fact that only \texttt{POST} can create resources, which is the prerequisite for executing \texttt{GET}, \texttt{PUT}, and \texttt{DELETE}. Also, \texttt{DELETE} and \texttt{PUT} may delete or alternate the resource, so we should create multiple resources via \texttt{POST} to benefit the successor nodes which may require the resource. We force \texttt{GET} to be the first test case because either a node (\textit{e.g.,} root node in Figure~\ref{fig:tree_hierarchy}) requires no parameter for launching a \texttt{GET} request or we could retrieve its required parameters from the parent node. Again, note that the order in our current template is ad hoc for fuzzing purposes, there could be other orders which might be more efficient.

Algorithm~\ref{alg:fuzzing} demonstrates our basic test case generation algorithm. The outer for loop iterates over the trees of an application. For each tree, we traverse the tree via a depth-first order and recover the endpoint URLs. For each method of an endpoint, we obtain its required parameters and retrieve values for each parameter from the resource pool. There are two key procedures in this process: \texttt{Search()} and \texttt{FuzzyMathing()} for searching candidate resources, and \texttt{Retrieve()} for retrieving resource values. \texttt{Search()} is based on the annotation table. It returns the unique resource identifier if such an entry exists in the annotation table, or returns the original name otherwise. \texttt{FuzzyMathing()} searches for candidate resources along the tree path backward with the heuristics discussed in Section~\ref{sec:fuzzymatching} until a candidate is found. Since there could be multiple matched resource identifiers, it selects a random one with matching score above a threshold. In order to further empower the fuzzing ability of our approach, we also enable \texttt{FuzzyMathing()} to return none no matter whether a matched resource is found or not. If \texttt{FuzzyMathing()} returns none, the \texttt{Retrieve()} function will use an arbitrary value generated by a simple fuzz engine, \textit{i.e.,} a random value of string, integer, or UTF-8 depending on the type of the parameter.

\subsubsection{Response Analysis and Resource Extraction}

We execute each test case immediately after it has been generated and analyze the response message. If the response code is 2XX, which implies a successful request, we can then extract the resources and save them into the corresponding node of the resource pool. Meanwhile, a successful request also implies the fuzzing pairs generated by \texttt{FuzzingMatching()} is useful, and we should increase the probability of using these pairs for composing new requests later via \texttt{IncreaseMatchingScore()}. If the response code is 4XX, which implies a bad request, we should decrease the probability of using the pairs via \texttt{DecreaseMatchingScore()}. Finally, a response code of 5XX implies a bug has been detected.

\begin{figure*}[t]
\centering
\begin{subfigure}{0.49\textwidth}
\includegraphics[width=\linewidth]{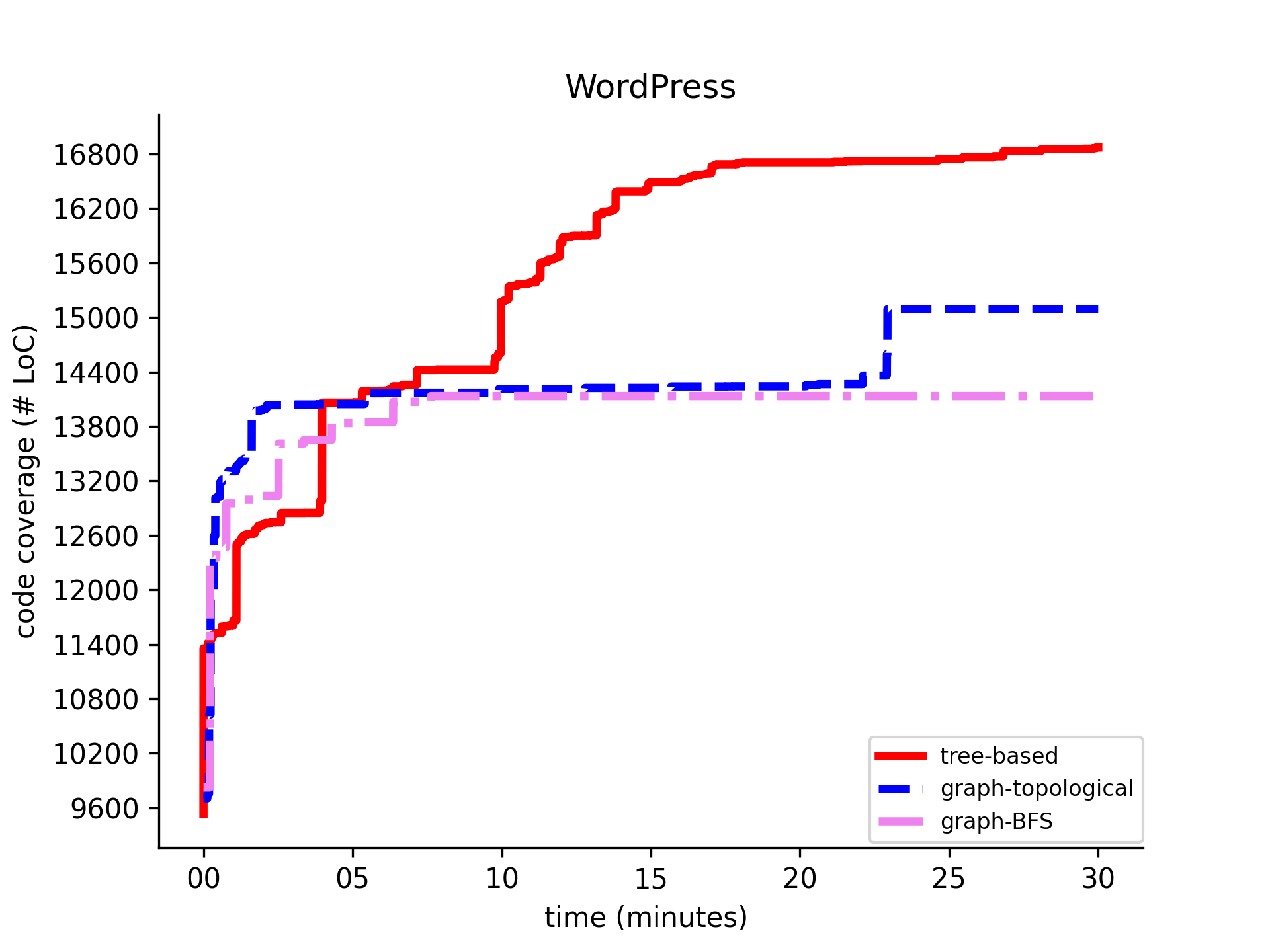}
\caption{Coverage for WordPress in 30 minutes.}
\label{fig:comp-tree-30min}
\end{subfigure}
\begin{subfigure}{0.49\textwidth}
\includegraphics[width=\linewidth]{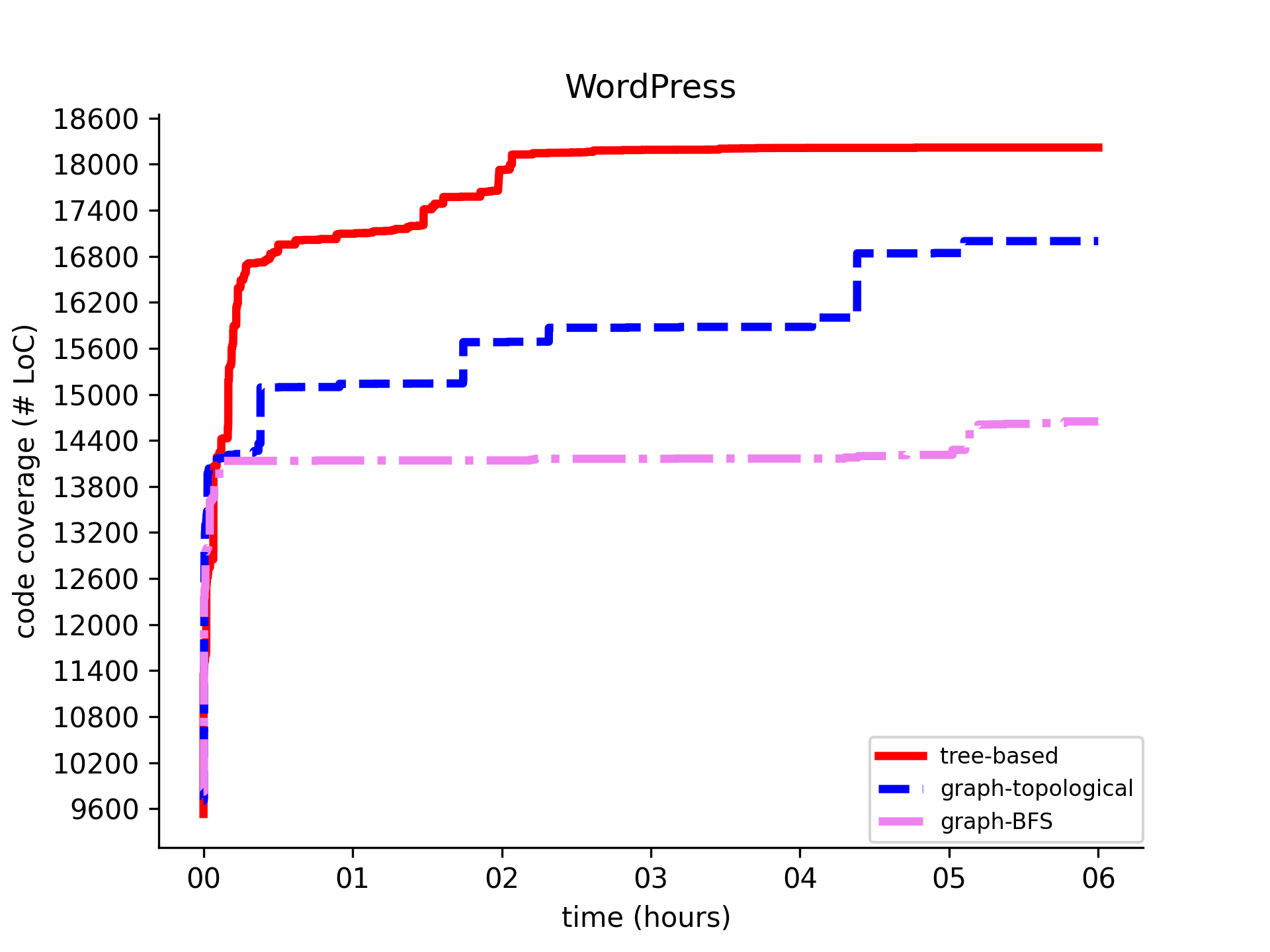}
\caption{Coverage for WordPress in 6 hours.}
\label{fig:comp-tree-6h}
\end{subfigure}
\caption{Comparing the efficiency of our tree-based approach with two graph-based approaches, bread-first search and topological sort.}
\label{fig:comp-tree}
\end{figure*}

\subsection{Overall Fuzzing Strategy}
Our basic method in Algorithm~\ref{alg:fuzzing} only considers required parameters. The number of generated test cases is linear to the number of APIs, \textit{i.e.,} $\leq (k+3) * API\#$. Therefore, it can finish execution in a short period. However, besides required parameters, an API may also support dozens of optional parameters. Generating test cases with optional parameters supplied should be useful for achieving better code coverage. 

Our overall fuzzing strategy is based on the Algorithm~\ref{alg:fuzzing} but considers both required and optional parameters. The idea is straightforward, \textit{i.e.,} we firstly employ the basic method for each API with no optional parameters, and gradually increase the number of optional parameters. For example, if an API supports $n$ optional parameters, we should generate $n$ different test cases, where each one specifies a unique optional parameter. Suppose $m$ of the $n$ requests are successful, then we generate $m*(m-1)$ test cases in the next round, where each test case contains two optional parameters selected from the successful ones. We continue this process gradually with more optional parameters until all the requests with a certain number of operational parameters fail. Note that exhausting all combinations ($n!$) of optional parameters for an API is impractical. Our approach can largely reduce the number of fruitless requests but should also be able to achieve a similar test effectiveness considering the number of dependencies (resources) can be successfully solved. 

Figure~\ref{fig:alg-demo} presents an toy example for better demonstrating our overall fuzzing strategy. Suppose the API tree in Figure~\ref{fig:alg-example} contains three nodes, \texttt{N1} and \texttt{N3} which are token nodes, and \texttt{:n2} which is a parameter node. We create two sub resource pools for \texttt{N1} and \texttt{N3}. Then we generate test cases for \texttt{N1}, \texttt{:n2}, and \texttt{N3} one by one. For each node, we first generate the test case of a \texttt{GET} request with required parameters only. Then we generate test cases with optional parameters for the API. Next, we generate $k$ test cases for \texttt{POST} requests with required parameters only, followed by such requests with optional parameters. We continue the test case generation process for \texttt{PUT} and \texttt{DELETE}. After all test cases for the API tree have been executed, we can continue the fuzzing process and restart from \texttt{N1}. Note that our second fuzzing round can generate new test cases different from the first round, \textit{e.g.,} due to the randomness of the \texttt{FuzzingMatching()} function and updated matching scores of dependency pairs by {IncreaseMatchingScore()} or \texttt{DecreaseMatchingScore()}.

\begin{table*}[]
\centering
\caption{Detailed experimental data of our comparison experiment with RESTler and EvoMaster in six hours. The LoC incerase is calculated as (LoC of foREST - LoC of EvoMaster)/LoC of EvoMaster.}
\small
\begin{tabular}{|c|c|c|ccc|ccc|ccc|c|}
\hline
\multirow{2}{*}{Project} &
  \multirow{2}{*}{API Group} &
  \multirow{2}{*}{\# APIs} &
  \multicolumn{3}{|c|}{RESTler} &
  \multicolumn{3}{|c|}{EvoMaster} &
  \multicolumn{4}{|c|}{foREST}
  \\ \cline{4-13} 
 &
   &
   &
  \# requests & LoC & \# bugs &
  \# requests & LoC & \# bugs &
  \# requests & LoC & \# bugs & LoC increase \\ \hline
  WordPress & all & 39 &
  6249 & 15111 & 0 &
  6972 & 16183 & 0 &
  7338 & 18217 & 3 & 12.5\%\\ \hline
\multirow{3}{*}{GitLab} &
  projects & 33 &
  281526 & 1838 & 2 &
  107297 & 7298 & 0 &
  90470 & 10151 & 3 & 39.1\%\\ \cline{2-13} 
 & groups & 23 &
  377780 & 1407 & 0 &
  89212 & 6668 & 2 &
  89865 & 7433 & 2 & 11.5\% \\ \cline{2-13} 
 &
  commits & 18 &
  375126 & 1900 & 0 & 
  107200 & 8406 & 0 &
  109533 & 15342 & 3 & 82.5\% \\ \hline
Total &
  - & 113 &
  1040681 & 20256 & 2 & 
  310681 & 38553 & 2 &
  297206 & 51143 & 11 & 32.7\% \\ \hline
\end{tabular}
\label{tab:result}
\end{table*}

\begin{figure*}[th]
\centering
\begin{subfigure}{0.49\textwidth}
\includegraphics[width=\linewidth]{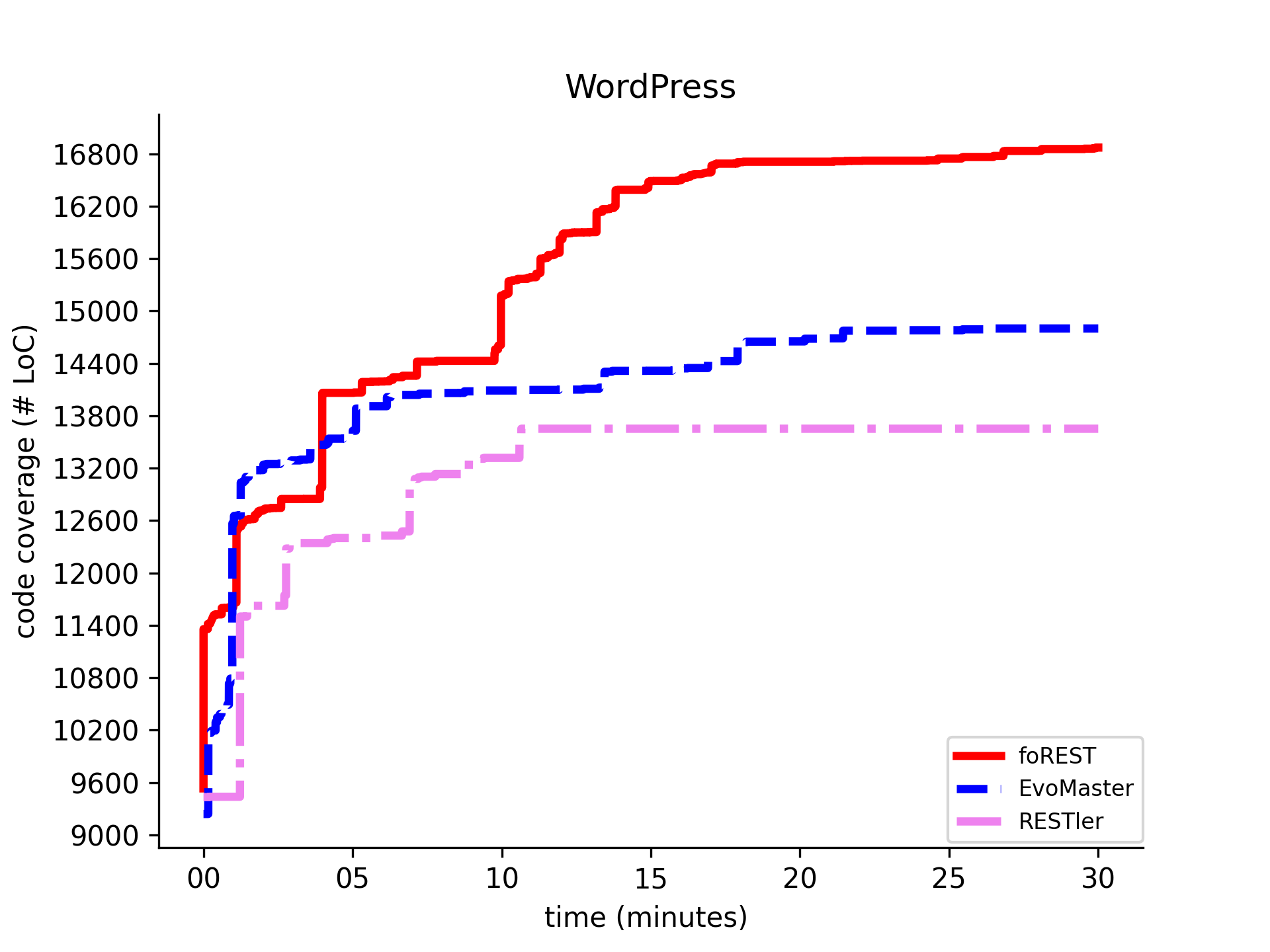}
\caption{Coverage for WordPress in 30 minutes.}
\label{fig:result-wordpress-30min}
\end{subfigure}
\begin{subfigure}{0.49\textwidth}
\includegraphics[width=\linewidth]{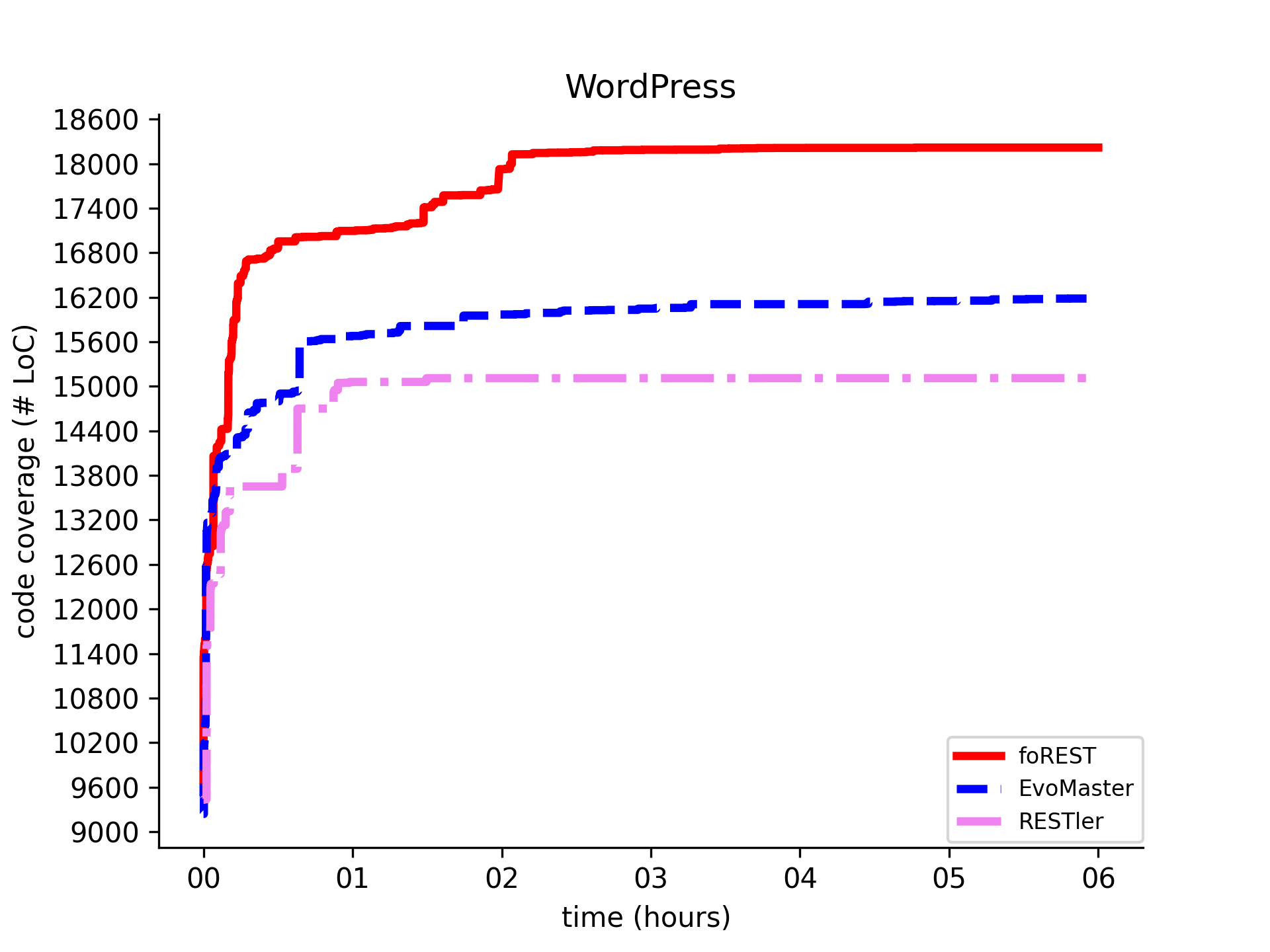}
\caption{Coverage for WordPress in 6 hours.}
\label{fig:result-wordpress-6h}
\end{subfigure}
\caption{Experimental result of comparing foREST with RESTler and EvoMaster in fuzzing WordPress.}
\label{fig:result-wordpress}
\end{figure*}

\section{Evaluation}\label{sec:evaluation}
This section presents our evaluation experiments based on a prototype implementation of foREST. In particular, we are interested in two questions. 

\begin{itemize}
\item \textbf{RQ1:} Compared to traditional approaches based on API dependency graphs, can our tree-based approach be more efficient in practice? 
\item \textbf{RQ2:} Will foREST perform better than other existing tools, such as RESTler~\cite{atlidakis2019restler} and EvoMaster~\cite{EvoMaster}?
\end{itemize}

Next, we present our experimental setting and experimental results to answer these two questions.

\subsection{Implementation and Experimental Setting}
We have implemented a prototype tool for the proposed tree-based fuzzing approach with python 3.8, including both the tree-based structure for API modeling and the corresponding resource management and retrieving strategies. It contains 2K+ lines of python code. The tool supports RESTful API specifications in either Swagger 2.0 or Open API standards. We release our tool as open source, and it is available online.

We conduct our evaluation experiment in a local network environment, \textit{i.e.,} we deploy foREST on a PC and test it against local RESTful services. Our experimental applications include two widely-employed open source RESTful applications (\textit{i.e.,} WordPress and GitLab). We normally test each application for six hours and examine the achieved code coverage. More experimental details are provided later in each sub experiment.

\subsection{Efficiency of Tree-based Approach}

In order to study the efficiency of our novel tree-based approach, we also implement another version with traditional graph-based approaches using the same foREST framework. For fair comparison, we implement two representative graph parsing algorithms, BFS and topological sort. The BFS algorithm directly employs the approach proposed in RESTler~\cite{atlidakis2019restler}; for topological sort, we employ a resource pool to buffer the result generated in previous request. Therefore, we have three configurations of fuzzing tools that only differ in the test case generation algorithms.

We fuzz WordPress for six hours with these three configurations and measure the covered lines of code. Our result is presented in Figure~\ref{fig:comp-tree}. Figure~\ref{fig:comp-tree-30min} demonstrates their coverage growths in the first 30 minutes with more observable differences. Each coverage upsurge on the graph generally implies a new API has been successfully triggered. We can observe that the three approaches perform comparably in the first ten minutes, and they all cover about 14000 lines of code. However, the code coverage of our tree-based approach becomes prominent after that. Figure~\ref{fig:comp-tree-6h} presents the coverage information of all the six hours. We can obverse that our tree-based approach finally has covered most lines of code (18217), which is better than 16998 achieved by topological sort, and 14649 achieved by BFS. 

\begin{figure*}[ht]
\centering
\begin{subfigure}{0.42\textwidth}
\includegraphics[width=\linewidth]{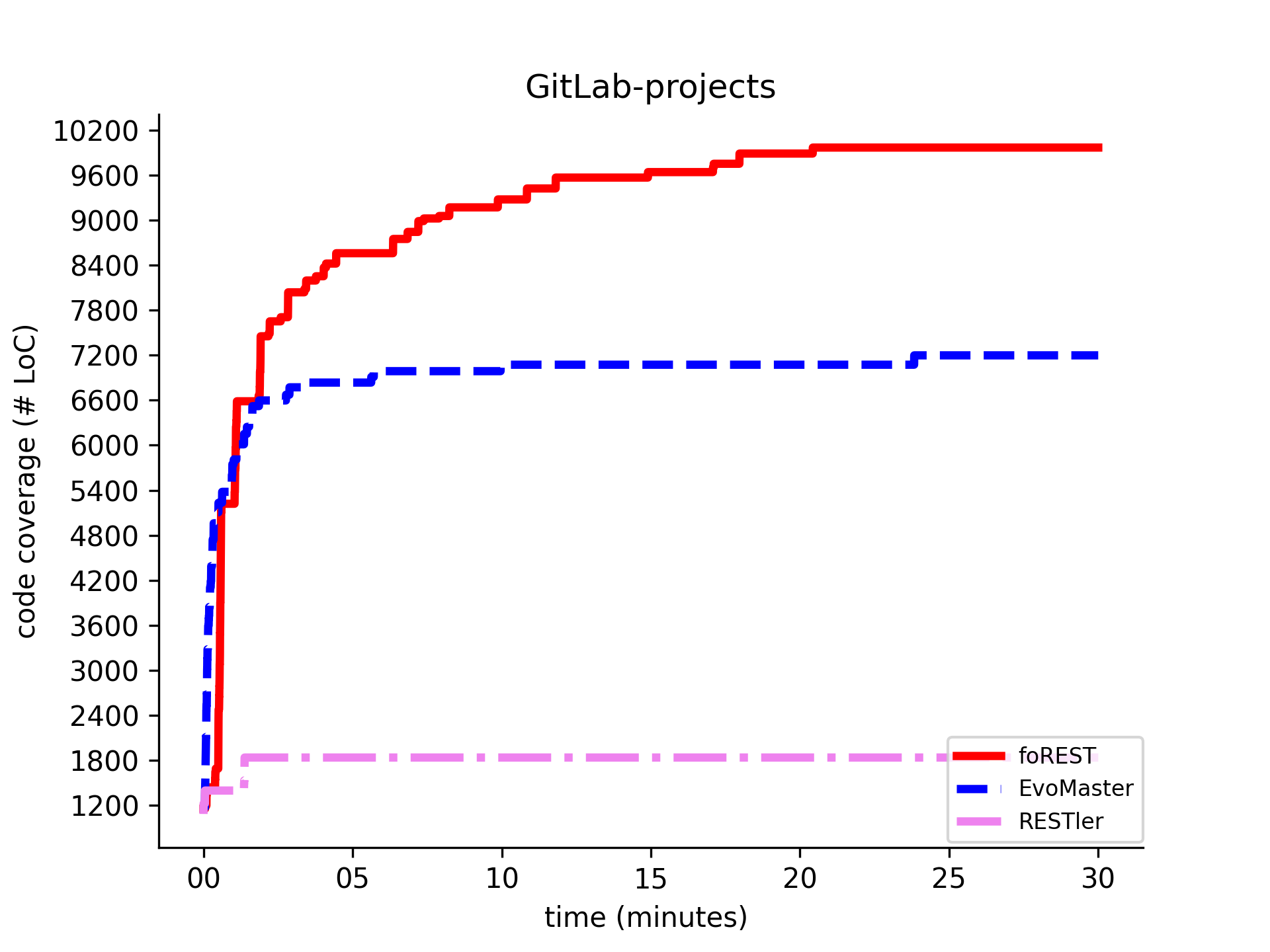}
\caption{Coverage for projects in 30 minutes.}
\label{fig:gitlab-projects-30min}
\end{subfigure}
\begin{subfigure}{0.42\textwidth}
\includegraphics[width=\linewidth]{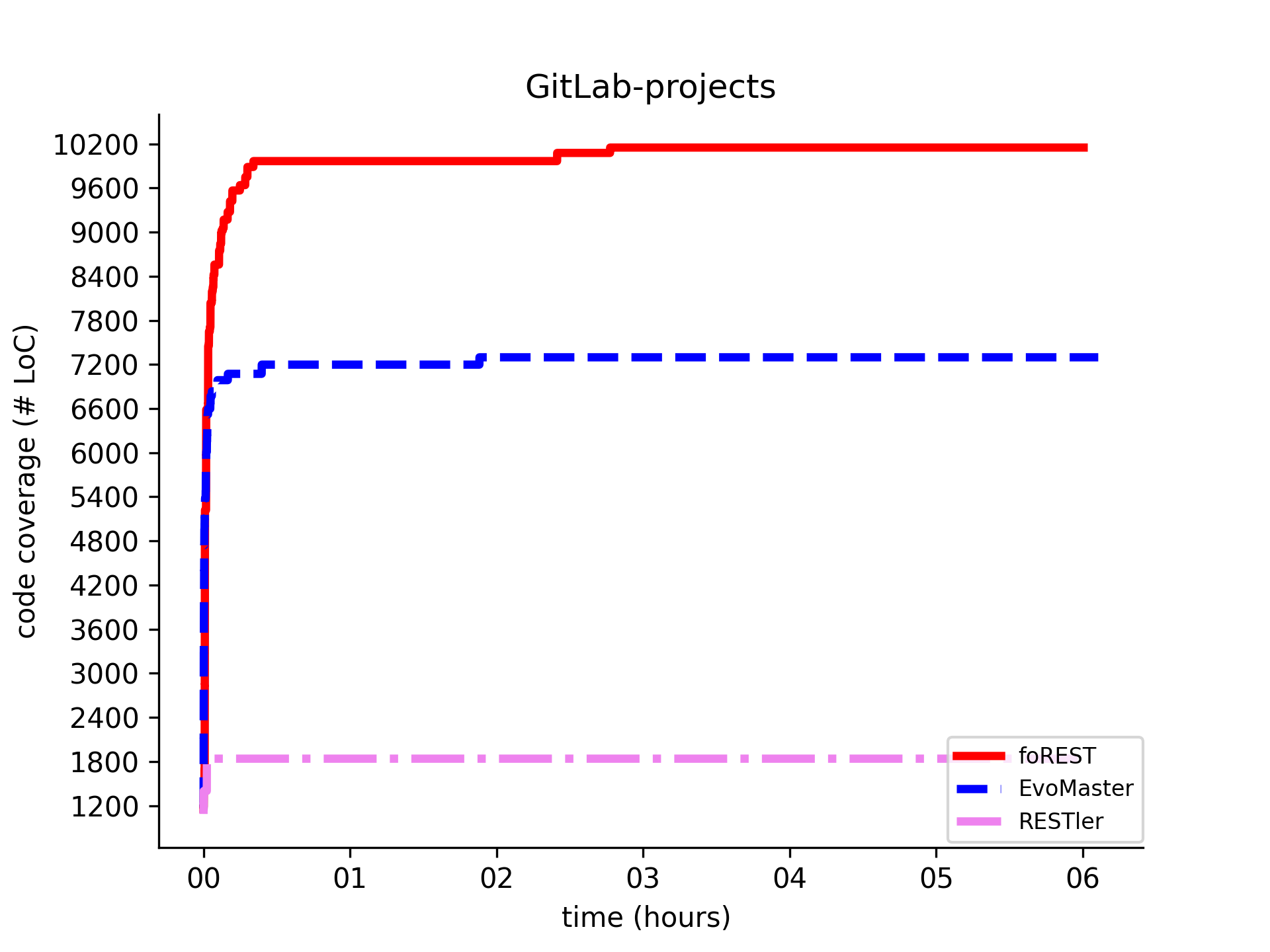}
\caption{Coverage for projects in six hours.}
\label{fig:gitlab-projects-6h}
\end{subfigure}

\begin{subfigure}{0.42\textwidth}
\includegraphics[width=\linewidth]{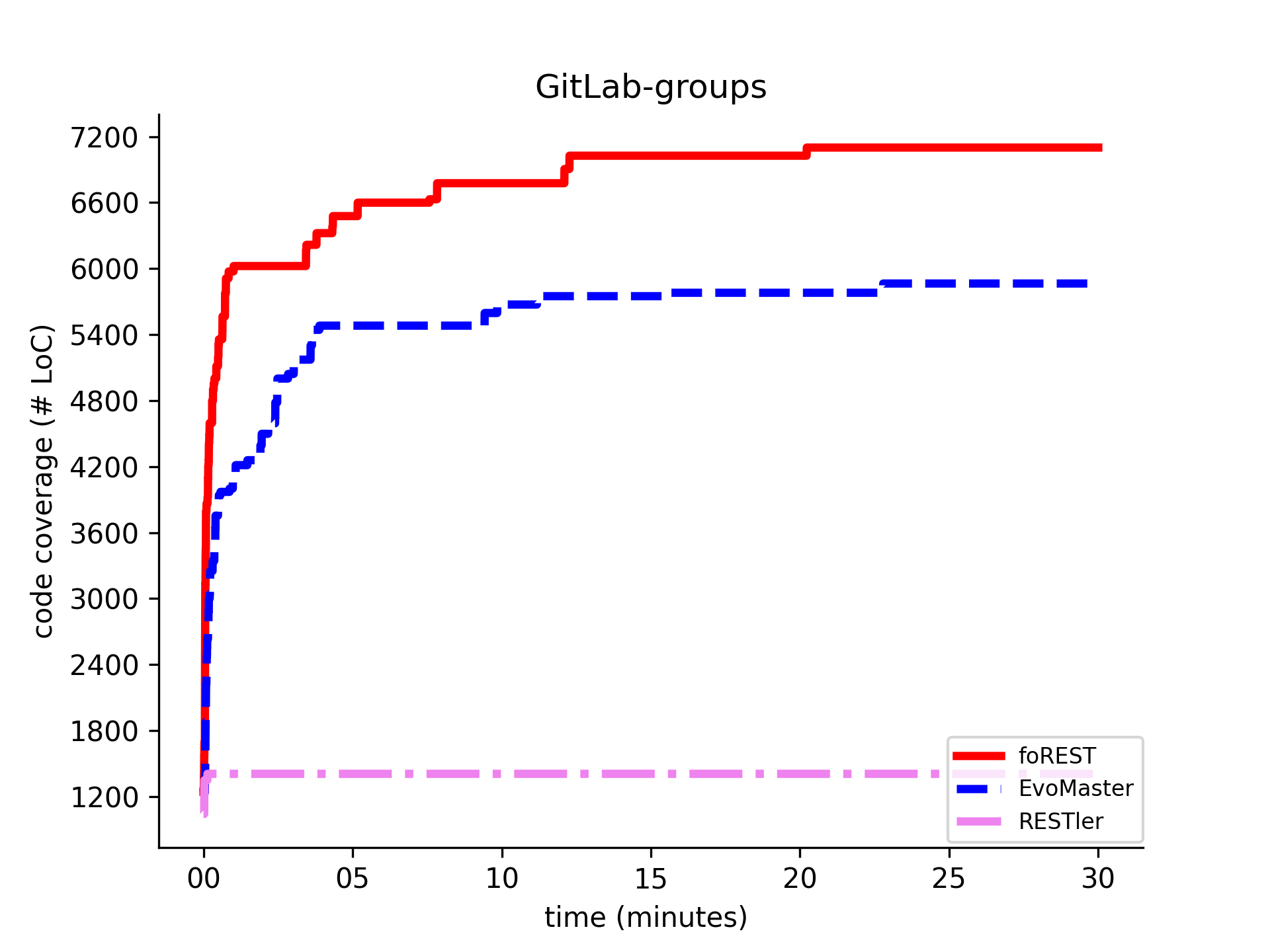}
\caption{Coverage for groups in 30 minutes.}
\label{fig:gitlab-groups-30min}
\end{subfigure}
\begin{subfigure}{0.42\textwidth}
\includegraphics[width=\linewidth]{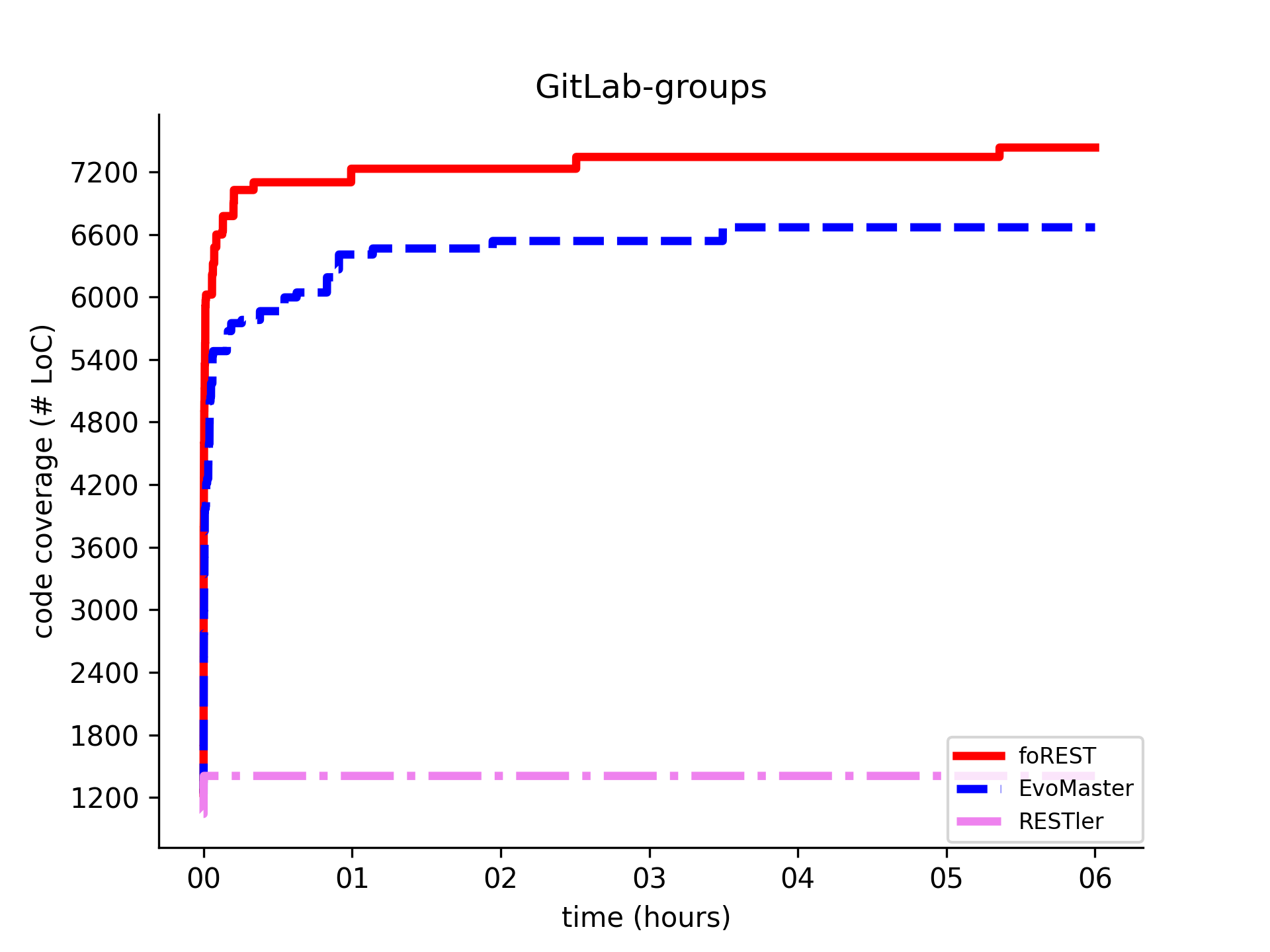}
\caption{Coverage for groups in six hours.}
\label{fig:gitlab-groups-6h}
\end{subfigure}

\begin{subfigure}{0.42\textwidth}
\includegraphics[width=\linewidth]{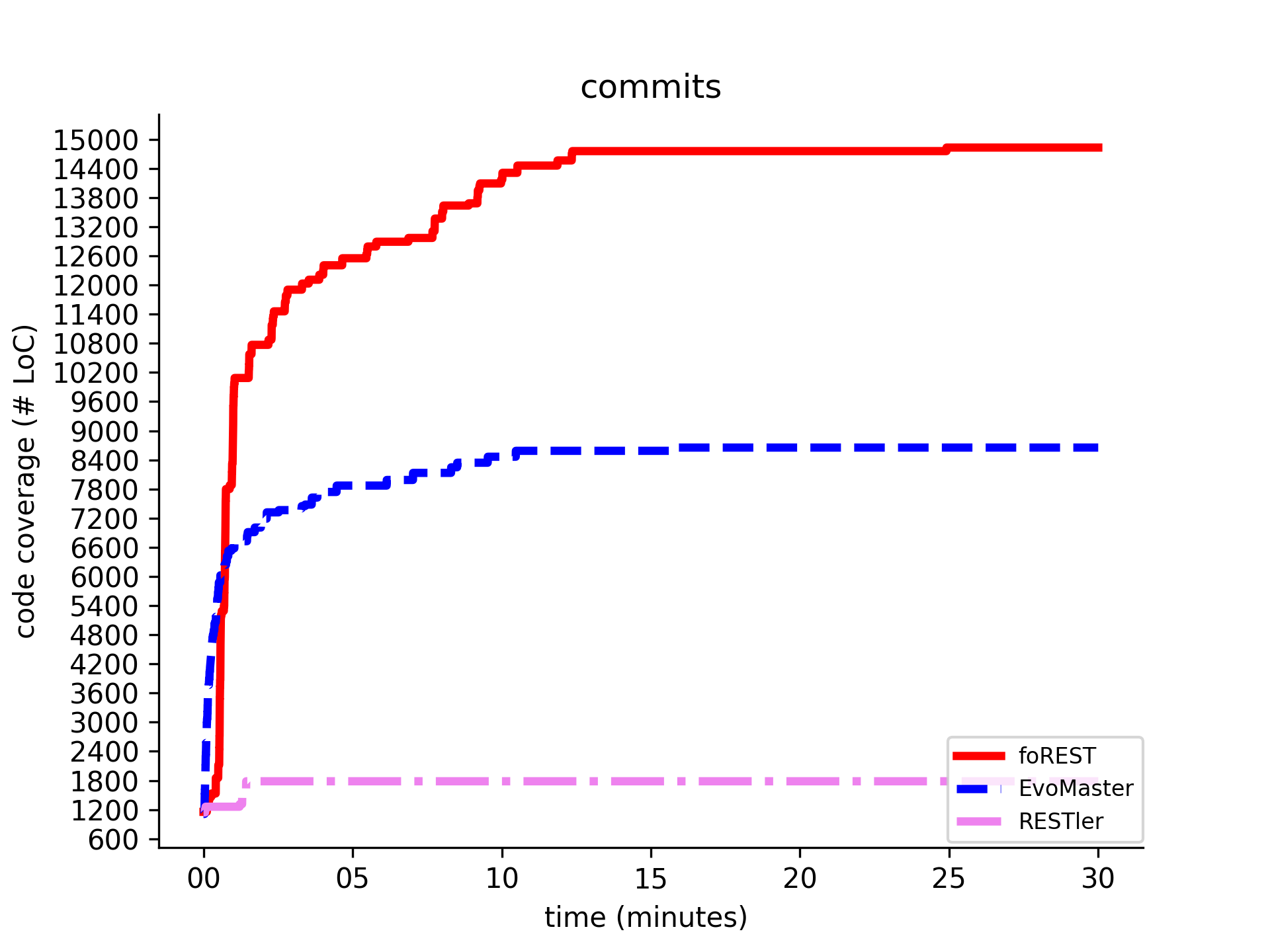}
\caption{Coverage for commits in 30 minutes.}
\label{fig:gitlab-commits-30min}
\end{subfigure}
 \begin{subfigure}{0.42\textwidth}
\includegraphics[width=\linewidth]{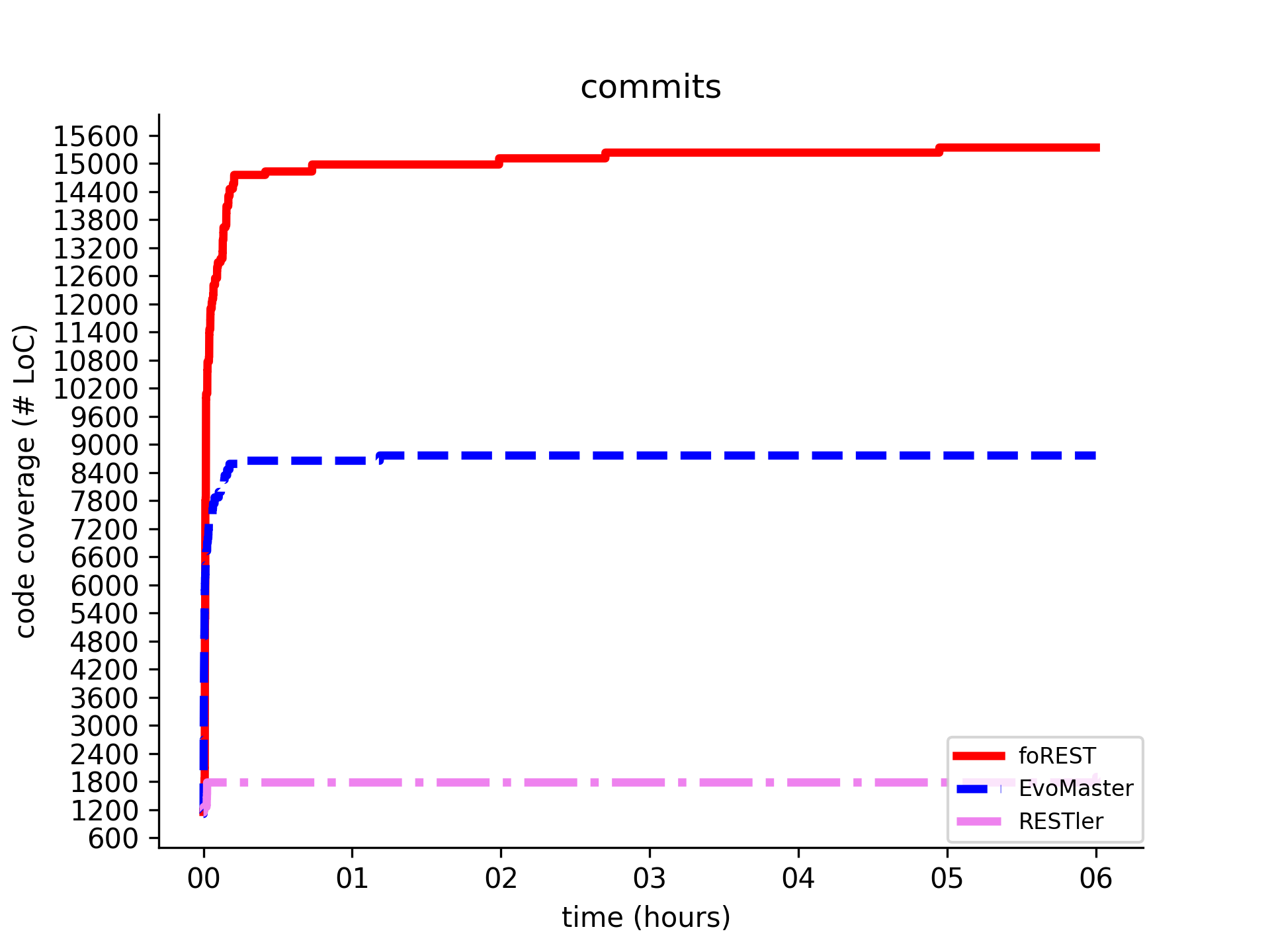}
\caption{Coverage for commits in six hours.}
\label{fig:gitlab-commits-6h}
\end{subfigure}
\caption{Experimental result of comparing foREST with RESTler and EvoMaster in fuzzing GitLab. GitLab has several API groups, we choose three API groups (projects, groups, and commits) and fuzz them seperately.}
\label{fig:result-gitlab}
\end{figure*}

Note that our experiment employs the same request frequencies for the three configurations, and their final request numbers are comparable. Therefore, we may conjecture that our tree-based approach is generally more efficient than traditional graph-based approaches. Meanwhile, it does not undermine the fuzzing effectiveness in code coverage. Our comparison experiments with RESTler in the next section can also coincide with this conjecture.

\subsection{Comparison with Existing Tools}
Now, we compare the performance of foREST with two state-of-the-art fuzzing tools for RESTful APIs, \textit{i.e.,} RESTler~\cite{atlidakis2019restler}, \cite{corradini2021empirical} and EvoMaster~\cite{EvoMaster}. RESTler is a graph-based black-box fuzzing tool collaborated by Columbia University and Microsoft. It adopts the same problem setting as we do, which only requires the API specification. EvoMaster is another powerful tool that supports both white-box and black-box mode. For fair comparison, we use its black-box mode (\texttt{--blackBox true}) with a request rate 300 per minute (\texttt{--ratePerMinute 300}) in our experiment.

We apply these three tools to fuzz two open source applications, WordPress and GitLab. Figure~\ref{fig:result-wordpress} shows our comparison result for WordPress in 30 minutes and six hours separately. We can observe that the coverage of foREST grows faster that RESTler and EvoMaster in the begining. Although EvoMaster outperforms foREST for a little while, foREST exceeds EvoMaster after the first five minutes, and the leading superiority becomes more obvious sooner after the first ten minutes. Table~\ref{tab:result} presents the detailed information after six hours. foREST has covered 18217 lines of code, with three new bugs found. EvoMaster covered 16183 lines of code with no bugs detected. RESTler covered 15111 lines of code with no bugs detected. The improvement of code coverage is about 12.5\%. Although foREST sends a bit more numbers of requests than other tools, the advantages of foREST is prominent. 

To better compare the effectiveness of our tool, we conduct another group of fuzzing experiments with GitLab. GitLab contains several groups of APIs\footnote{https://docs.gitlab.com/ee/api/api\_resources.html}. We choose three widely-employed groups to conduct of our fuzzing experiments (\texttt{projects},\texttt{groups}, and \texttt{commits}) and report the performance on these API groups separately. Our results are presented in Figure~\ref{fig:result-gitlab}. In general, all the results are consistent with WordPress. foREST performs the best among the three tools, and EvoMaster is better than RESTler. According to the detailed statistics in Table~\ref{tab:result}, RESTler has sent more requests than foREST and EvoMaster in six hours, and foREST has sent a similar number of requests as EvoMaster. Note that such speed could be affected by the response rate of the service. Finally, foREST has covered 10151 lines of code for the API group of projects and has detected three new bugs. EvoMaster covered 7298 lines of code with no bugs found. The improvement is about 39.1\%. Similarly, the improvement of foREST for the API group of groups and commits are 11.5\% and 82.5\% respectively. foREST has detected eight new bugs for GitLab in total, while the other tools detected none.

Based on the previous results, we can conclude that foREST has non-trivial performance advantages over RESTler and EvoMaster in fuzzing RESTful APIs. Note that that are also other tools that support RESTful API fuzzing, such as RestTestGen~\cite{viglianisi2020resttestgen} and bBOXRT~\cite{laranjeiro2021black}. However, an empirical study~\cite{corradini2021empirical} has shown that these tools are either preliminary or not robust compared to RESTler, and RESTler is the best one over them. 

\section{Related Work}\label{sec:related}
As REST surges into popularity, there are dozens of papers working on RESTful API testing. A large portion of these work investigates on how to define test-oriented specifications, including \cite{chakrabarti2009test,pinheiro2013model,lamela2013towards,fertig2015model,segura2017metamorphic,karlsson2020quickrest,karlsson2021automatic}. Similar to traditional software testing, one major problem for RESTful test case specification lies in how to provide an oracle for test case verification, such as via property-based~\cite{pinheiro2013model,karlsson2020quickrest,karlsson2021automatic} and metamorphic testing~\cite{segura2017metamorphic}. Such problems and investigations are not directly related to our work.

Our work focuses on RESTful API fuzzing, and there are several papers (\textit{e.g.,}~\cite{ed2018automatic,atlidakis2019restler,atlidakis2020pythia,EvoMaster,arcuri2020automated,viglianisi2020resttestgen,corradini2021empirical}) work on the same problem. As we have studied in this paper, the critical problem for RESTful API fuzzing lies in how to model the dependency relationships for generating valid requests. Existing representative work that studies the problem and supports stateful fuzzing includes RESTler~\cite{atlidakis2019restler}, EvoMaster~\cite{EvoMaster,arcuri2020automated}, and RESTTESTGEN\cite{viglianisi2020resttestgen}. Next, we mainly compare our work with these three tools.

RESTler~\cite{atlidakis2019restler} is a stateful RESTful API fuzzing tool. It employs a graph-based approach to model the dependencies among APIs by analyzing API specifications. Each test case generated by RESTler is a sequence of requests in order to be stateful, \textit{i.e.,} the front requests aim to arrive at the state required by the last request. It traverses the API dependency graph to generate such sequences. RESTler is a collaborative work with Microsoft, and there are several follow-ups, such as to enhance the fuzzing ability with neural networks ~\cite{atlidakis2020pythia}, to check security property violations~\cite{atlidakis2020checking}, and to perform regression testing with RESTler~\cite{godefroid2020differential}.

RestTestGen~\cite{viglianisi2020resttestgen} is another tool similar to RESTler. It also models API dependencies as graphs but should be more powerful in the capability of dependence inference. Instead of strict string matching, RestTestGen employs several fuzzy matching strategies, such as case insensitive and stemming. Furthermore, it also introduces abnormal case generation strategies, such as mutation, missing required parameters, or incorrect input type. Corradini \textit{et al.}~\cite{corradini2021empirical} have conducted an experiment to compare the performance of RestTestGen with RESTler, as well as two other tools, bBOXRT~\cite{laranjeiro2021black} and RestTest~\cite{martin2020restest}, and their result shows that RESTler performs the best in the benchmark and it is the most robust tool.

EvoMaster~\cite{EvoMaster,arcuri2020automated} is a white-box testing tool but also supports black-box mode. Similar to our work, it has no API dependency graph. In order to generate a valid test case for a particular endpoint, it heuristically searches the antecedent requests or preconditions that should be fulfilled along the path of the endpoint URL. EvoMaster is a state-of-the-art tool, and has attracted follow-up work, such as \cite{zhang2019resource} that enhances the verification ability of EvoMaster for particular RESTful services.

Besides inferring the dependencies based on the default API specification. There are also other papers that resorts to extra input, such as manual dependency specification~\cite{martin2019catalogue,martin2021specification}, or historical traffic data with machine learning~\cite{mirabella2021deep}. These approaches are orthogonal to other work. 

To summarize, our work tackles the RESTful API fuzzing problem, which has also been studied by RESTler~\cite{atlidakis2019restler}, RestTestGen~\cite{viglianisi2020resttestgen}, and EvoMaster~\cite{EvoMaster,arcuri2020automated}. The novelty of our paper lies in that it serves as the first attempt to propose a systematic tree-based approach for RESTful API fuzzing. Our approach has been shown more efficient leveraging several subtle design choices, such as tree-based API modeling and resource management, fuzzing matching and dynamic dependency pair evaluation strategies, \textit{etc}. 

\section{Conclusion}\label{sec:conclusion}
This work has studied the problem of black-box RESTful API fuzzing based on the API specifications in Swagger or OpenAPI formats. The challenge mainly lies in how to model the relationships of different APIs in order to solve resource dependencies and generate valid requests. To tackle this problem, this work proposes a novel tree-based approach to effectively capture the relationships among APIs. On one hand, it can largely simplify the dependency relationships among APIs employed by traditional graph-based approaches. On the other hand, the resource dependencies modeled by our approach are more accurate. To study the performance of our approach, we have implemented a prototype and conducted several groups of comparison experiments with widely-employed real-world REST services. Our experimental results first verified the efficiency of the tree-based approach compared to traditional graph-based approaches, and then showed that foREST can achieve a higher code coverage than state-of-the-art tools. We believe our proposed approach and prototype would be useful to the community in advancing the development of RESTful API fuzzing.

\bibliographystyle{ACM-Reference-Format}
\bibliography{forest}
\end{document}